# EXPERIMENTAL SUMMARY

Melvyn J. Shochet
Department of Physics and Enrico Fermi Institute
University of Chicago
Chicago, IL 60637 USA

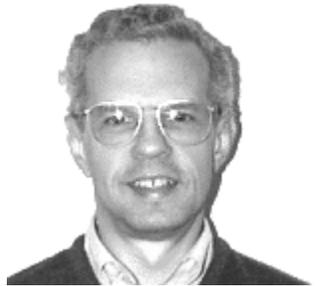

An overview is presented of the experimental talks given at the XXXVIIth Rencontres de Moriond QCD session.

# 1. INTRODUCTION

A great many experimental results were presented at the XXXVIIth Rencontres de Moriond QCD meeting. Unfortunately time and space limitations prevent me from summarizing them all. In particular, I will not discuss any future experiments; I expect we will be hearing much about them in the next few years. In the following sections, I will summarize heavy flavor physics, jet production, event shapes and correlations, structure functions, diffraction, non-QCD new physics, and heavy ion collisions. Names in brackets refer to the speakers, whose papers appear in this volume.

# 2. HEAVY FLAVOR PHYSICS

QCD plays an essential role in the study of the charm, bottom, and top quarks. The charm sector continues to be a test bed for understanding QCD. In the past, $c\bar{c}$ bound states were studied to probe the strong interaction at intermediate distance scales. In the future, charm decays will provide important tests of modern non-perturbative techniques. For the *b*-physics goal of understanding the origin of CP violation, QCD both helps, by binding the quarks into the neutral mesons, and hinders through the complications of Penguin diagrams. In top quark studies, QCD presents experimental problems. The effects of hard gluon bremsstrahlung produce the dominant systematic uncertainty in the top quark mass measurement.

The major change in the past year is the quantity of data that is becoming available for analysis. The B factories at KEK and SLAC have produced record luminosities much more quickly than many had expected. In CP violation, both BaBar and Belle presented new results. Figure 1 shows the BaBar $B \to J/\psi K^0$ asymmetries for $K_L$ and $K_S$ [Long]. The average $\sin 2\beta$ from Belle [Higuchi] and BaBar, assuming fully correlated systematic uncertainties, is $0.78 \pm 0.08$.

The first results on CP asymmetry in $B \to \pi^+\pi^-$ were presented. Figure 2 shows that the Belle asymmetry amplitudes, including the direct CP violating cosine term, are maximal [Sagawa]. The BaBar values, on the other hand, are close to zero [Long]. More statistics are clearly needed to resolve this important issue.

The large statistical samples allow a variety of other measurements that are important for fully understanding the CKM matrix. The values of $|V_{ub}|$ and $|V_{cb}|$ are being measured in a number of channels [Ishino,Kowalewski,Pappas,Barker]. Now that CLEO has seen that there aren't large new physics contributions to $b \to s\gamma$, they are using it to extract QCD parameters needed in other decays [Pappas]. New channels have been added at LEP and SLC to increase the sensitivity to $B_s$ mixing, with the current limit, $\Delta m_s > 14.9\,\mathrm{ps}^{-1}$ @ 95%CL [Abbaneo], approaching the favored Standard Model region. Finally, there are hints of rare decays, including $B \to \pi^0\pi^0$ at Belle [Gershon] and $B \to D_s^+\pi^-$ at BaBar [Fabozzi].

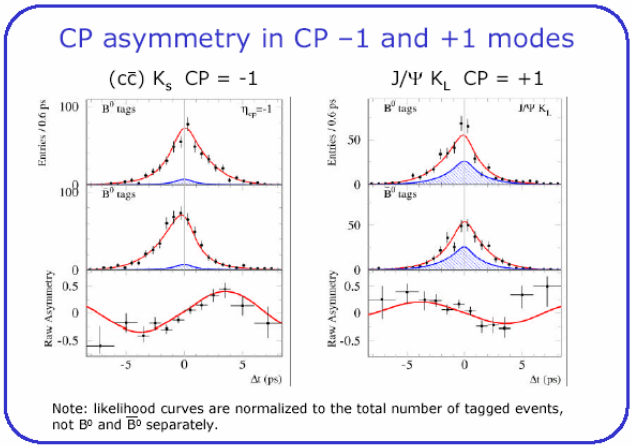 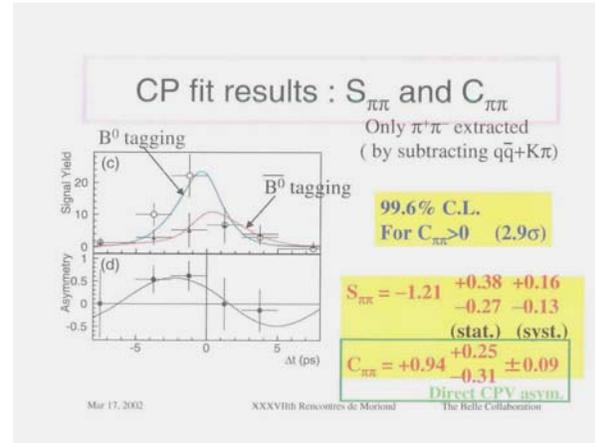

Fig. 1: BaBar $B \to J/\psi K^0$ asymmetries.  Fig. 2: Belle $B \to \pi^+\pi^-$ asymmetries.

The large data samples are also starting to provide answers to long standing questions in non-CKM B physics. An example is the relative lifetimes of the various B hadrons (Fig. 3), which show the expected hierarchy although perhaps with a range even larger than expected [Barker]. Tevatron run II results should reduce the uncertainties significantly and provide a definitive test.

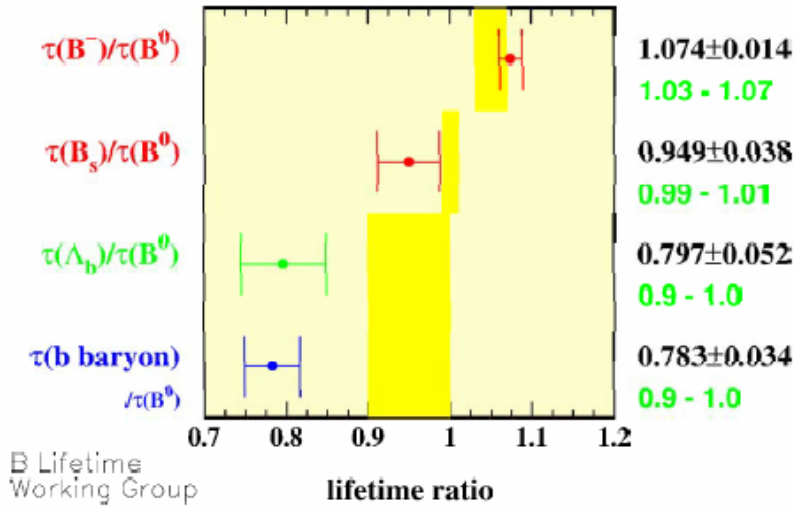

Fig. 3: Ratios of b-hadron lifetimes: data points and theory bands.

There were also new results presented in charm physics. E791 [Meadows] finds that to adequately describe $D^+$ Dalitz plots, additional scalar resonances are needed – a broad κ for the $K\pi\pi$ final state, and a narrow σ for $\pi\pi\pi$. BES [Haiping] has new high statistics results on the $\psi$ and $\psi'$. The $f_0(1710)$ is clearly seen in the $\gamma KK$ glueball search. They also have a new precision result on the mass of the $\eta_c$, $2977.6 \pm 0.8\ MeV$. CLEO is searching for $D^0\bar{D}^0$ mixing using $D^0 \to K_S\pi^+\pi^-$, which allows a determination of the strong phase [Smith]. They fit the Dalitz plot to many resonances and see for the first time a wrong-sign contribution (2σ in fit fraction, 3-4.5σ in amplitude).

In the near future, there should be much larger data sets. For charm, CLEO-c will provide serious tests of lattice QCD. In the $b$ system, BaBar and Belle will test CKM CP violation in many channels, while CDF and D0 will measure $B_s$ mixing. And from the Tevatron, we should see a precision top quark mass measurement as well as determinations of the top lifetime and couplings.

## 3. Hard QCD Scattering Processes

New comparisons of perturbative QCD to data have been carried out at HERA, LEP, and the Tevatron. For the latter, CDF has followed a suggestion of W. Giele to use the inclusive high-$P_T$ jet cross section at each $P_T$ and next-to-leading-order QCD to determine $\alpha_S$ [Seidel]. Figure 4 shows the running of $\alpha_S$ in good agreement with the world average $\alpha_S(M_Z)$, at least up to very high $P_T$ where the well known issue of the structure functions becomes important.

The LEP experiments have determined $\alpha_S$ from the $Z \rightarrow$ 4-jet rate and angular correlations. The final ALEPH result, shown in Figure 5, gives $\alpha_S$ in good agreement with the world average [Bravo i Gallart].

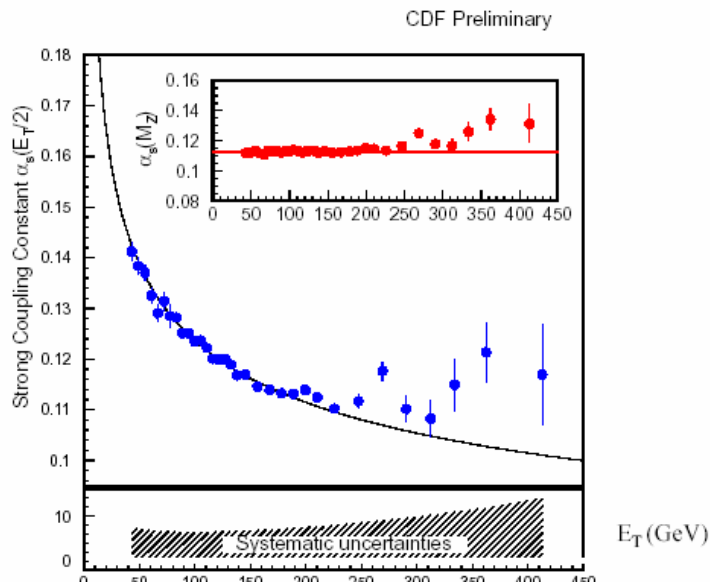
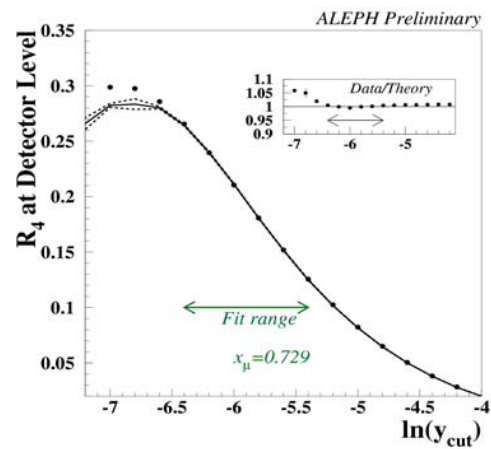

Fig. 4: The running of $\alpha_S$ using CDF data and NLO QCD.

Fig. 5: The rate of Z decay into 4 jets vs. clustering cutoff.

It has been known for a number of years that the $b$ production cross section at the Tevatron Collider is a factor of two larger than the NLO QCD prediction. At this meeting, we heard reports on $b$ production at HERA [Longhin] and in $\gamma\gamma$ collisions at LEP [Wengler]. In both cases, the measured cross section is also significantly larger than the theory prediction; at HERA the ratio is approximately 3.

**4. Event Shapes and Correlations**

Event shape parameters for hadronic jet final states are sensitive to the strong interaction coupling constant. In both $ep$ and $e^+e^-$ collisions (Fig. 6), the value of $\alpha_S$ is consistent with the expected running and the value measured at the Z pole [Movilla Fernandez, Rodrigues].

Correlated parton data provide additional information beyond that of the individual parton distribution functions. A number of such results were presented at this meeting. LEP studies of $\gamma\gamma \to p\bar{p}$ [Braccini] indicate that the proton has a diquark-quark rather than a 3-quark structure. In $\gamma\gamma \to$ jets at LEP [Wengler] and photoproduction at HERA [Jőnsson], there is better agreement with QCD when photon structure and multiple interactions are included. In WW production at LEP [Rabbertz], particle production in the region between jets of different W's compared to jets from the same W is providing information about the large color recombination systematic uncertainty in the W mass measurement.

Correlations among identical particles inform us about the underlying production dynamics. At LEP [Kress], pion pairs show the expected Bose-Einstein excess at small relative Q, while baryon pairs exhibit the Fermi-Dirac depletion. The source sizes show that $R_\pi > R_K > R_p$.

**5. Structure Functions**

A status report was presented on the first RHIC polarized proton operation [Bunce], whose goal is solving the proton spin problem. For existing data, there is a new HERA $F_2$ result [Zomer] that is shown in Figure 7. Once again, $\alpha_S$ agrees with the world average. The overall agreement among all the $\alpha_S$ measurements is one of the best indications of our understanding of QCD.

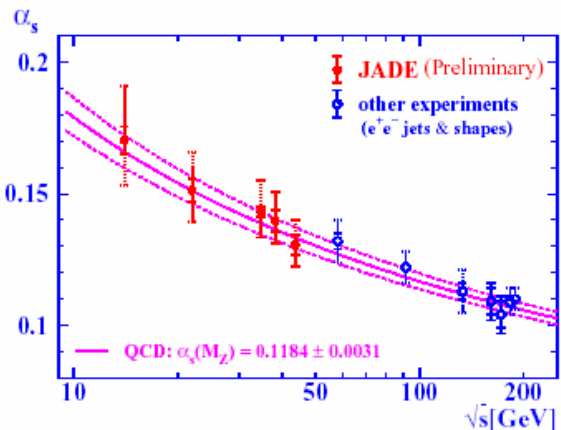

Fig. 6: Strong coupling from $e^+e^-$ jet-event shape variables.

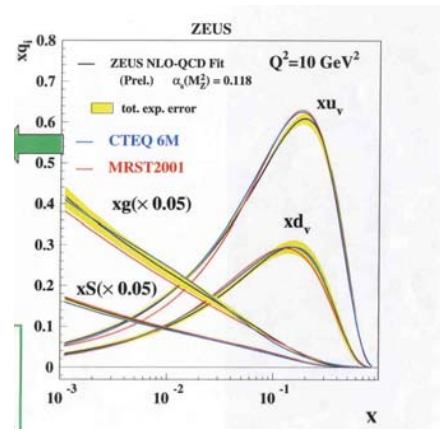

Fig. 7: New $F_2$ results from HERA.

## 6. Diffraction

The nature of diffractive scattering is being pursued on many fronts. At HERA, the study of deeply virtual Compton scattering (Figure 8) shows good agreement with QCD predictions [Grabowska-Bold]. More statistics are needed in order to test the *t* dependence of the cross section. Diffractive vector meson production has also been studied at HERA [Mohrdieck]. Figure 9 shows the power-law *t* dependence of the J/$\psi$ production cross section in good agreement with the QCD prediction. HERA inclusive diffraction data have properties expected if there is a large QCD gluonic component to the Pomeron: $F_2^D$ scaling violations (Fig. 10), the lack of a shrinking diffraction peak, and the observed $W$ and $Q^2$ dependence [Ruspa].

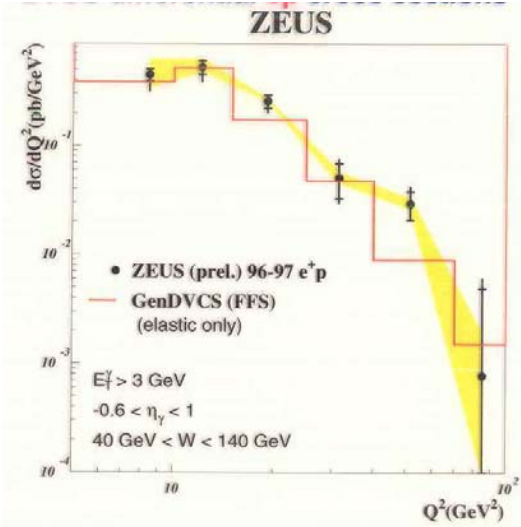
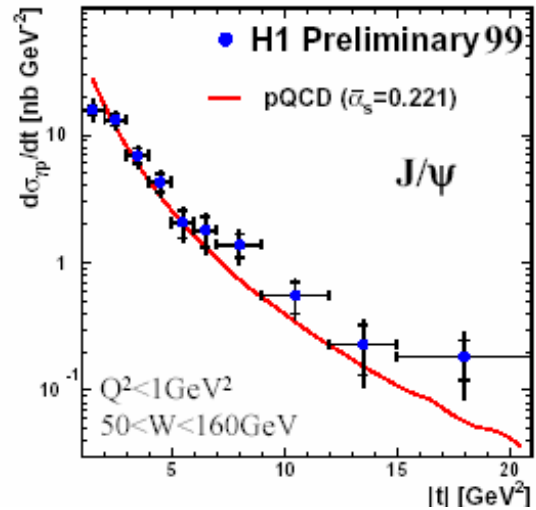

Fig. 8: Deeply virtual Compton scattering cross section as a function of $Q^2$ compared to QCD predictions.

Fig 9: Power-law *t* dependence of J/$\psi$ production at HERA.

At the Tevatron, CDF has results on multigap production, with gaps between the central region and both the projectile and target [Goulianos]. A model in which the gap probability is renormalized to one gives consistent explanations of single and double diffraction and double Pomeron exchange (Fig. 11).

In related phenomena, rapidity gaps have been studied in 3-jet Z decay at L3 [Field]. They place an upper limit of 7-9% on a color-singlet contribution, which is still consistent with previous Tevatron results. E791 data on diffractive dissociation of the $\pi$ have been used to understand the pion wavefunction [Ashery].

## 7. Non-QCD New Physics

The NuTeV collaboration has published a new precision measurement of $\sin^2\theta_W$ in neutrino deep-inelastic scattering [Bolton]. The use of almost pure neutrino and antineutrino beams allowed them to minimize the impact of the uncertain charm quark mass and thus significantly reduce the overall systematic uncertainty. Their result, $\sin^2 \theta_W = 0.2277 \pm 0.0013 \pm 0.0009$, is 3σ away from the world average (Fig. 12).

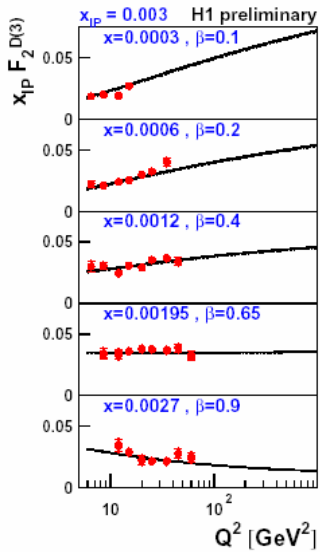

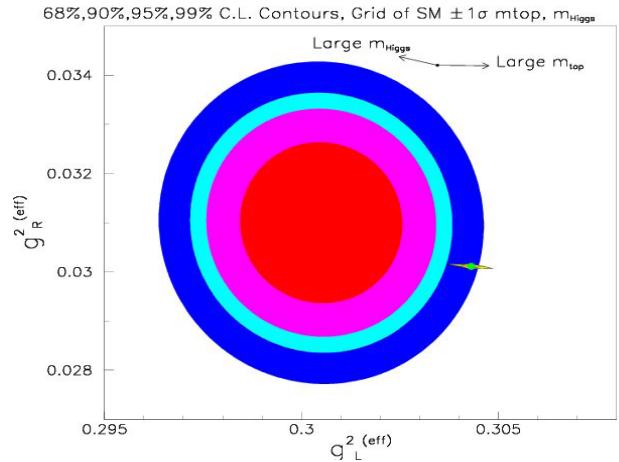

Fig. 10: Diffractive structure functions from HERA.

Fig. 12: The NuTeV $g_L$ and $g_R$ result compared to the world average.

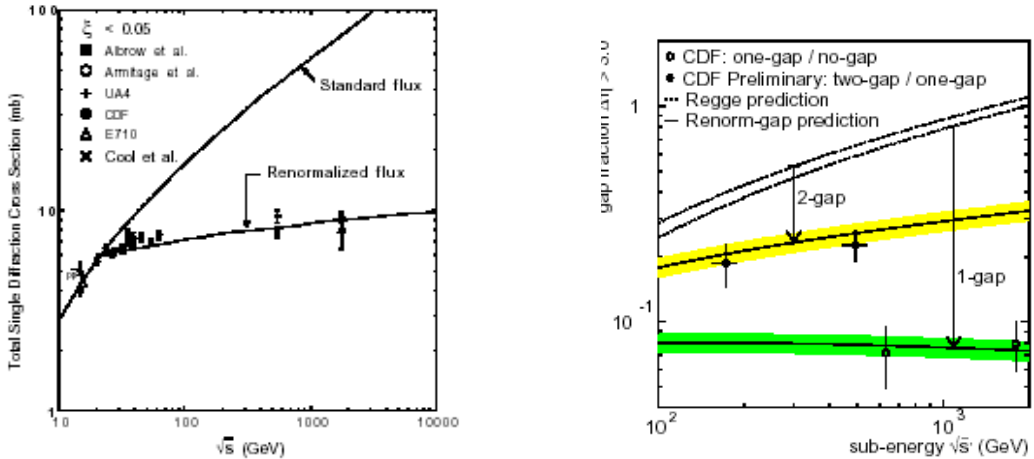

Fig. 11: Diffraction results from CDF. The left plot shows the single-diffractive cross section compared to the usual and renormalized Pomeron fluxes. The right plot shows the one-gap/no-gap and two-gap/one-gap ratios.

At LEP, the final results from the Higgs searches are mostly complete [Schwickerath]. The final ALEPH result retains the 3σ excess at 115 GeV. However each of the other experiments has at least a 20% probability that its observation in that region is a background fluctuation. All four LEP experiments have placed limits on extra dimensions by looking for $e^+e^- \rightarrow \gamma + \slashed{E}_T$, where the missing $E_T$ comes from graviton emission [Ganis]. Typical results are shown in Figure 13. Each experiment sets lower limits on $M_S$ varying from approximately 0.5 TeV for 6 large extra dimensions to almost 1.5 TeV for 2 large extra dimensions. ALEPH also searched for the stop and set an absolute lower limit on the stop mass of 65 GeV (Fig. 14) [Ganis].

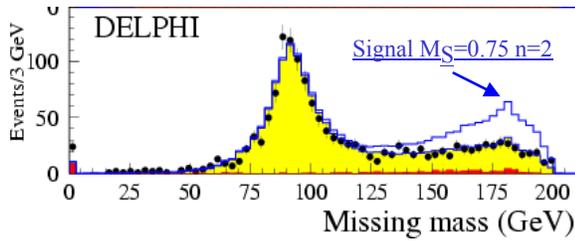
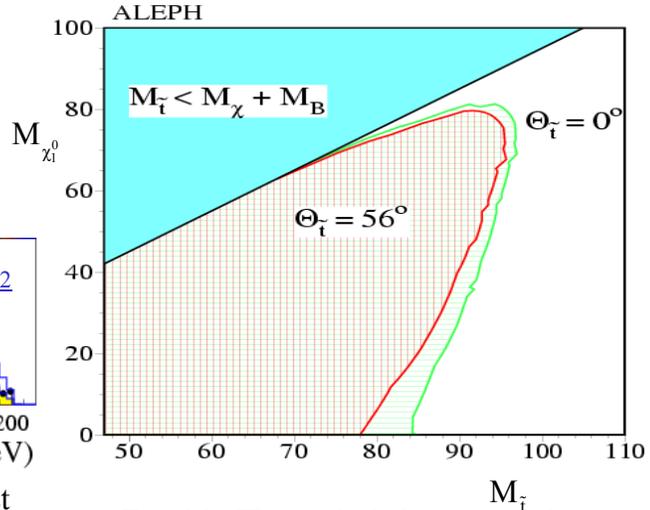

Fig. 13: Missing mass recoiling against a photon along with the signal expected from large extra dimensions.

Fig. 14: The excluded region in the search for the stop squark.

Large extra dimensions are also the focus of Tevatron searches [Muanza]. The D0 limits on graviton production, using monojets plus missing $E_T$ (Fig. 15), exceed those from LEP if the number of large extra dimensions is greater than 5. CDF looked for effects from a graviton propagator by searching for anomalous diphoton production (Fig. 16). They set a lower limit on the mass scale of 1 TeV.

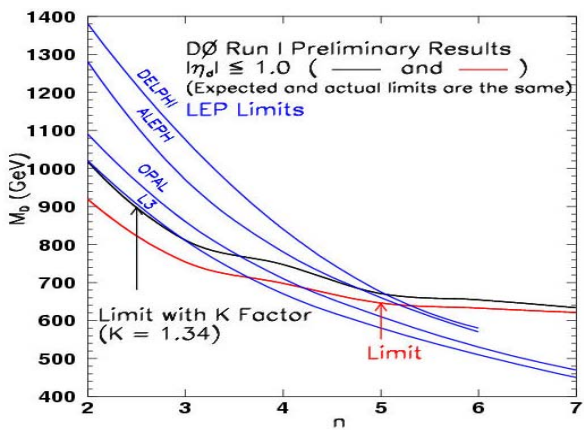
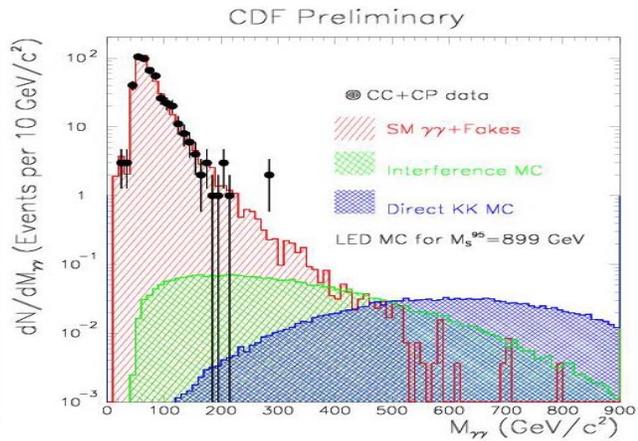

Fig. 15: Large extra dimension limits from the D0 monojet analysis.

Fig. 16: CDF search for large extra dimensions in diphoton data.

## 8. Heavy Ion Collisions

The first large statistics data samples from the RHIC experiments were presented at this meeting. The analyses are beginning to probe the physics of the high-energy, high-density QCD phase diagram, with the goal of investigating the quark-gluon plasma. The multiplicity in these events is mind boggling, with approximately 5000 charged particles in central collisions [Nouicer]. Figure 17 shows the negative charged particle multiplicity per unit rapidity [Xu]. Figure 18 gives the energy dependence of the multiplicity per unit rapidity for both gold-gold and proton-proton collisions [Rami]. Model discrimination will improve as more data is recorded.

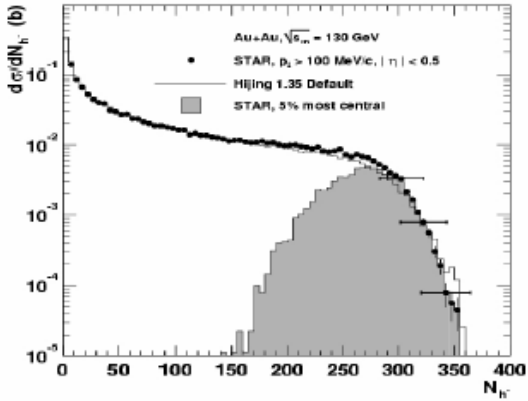
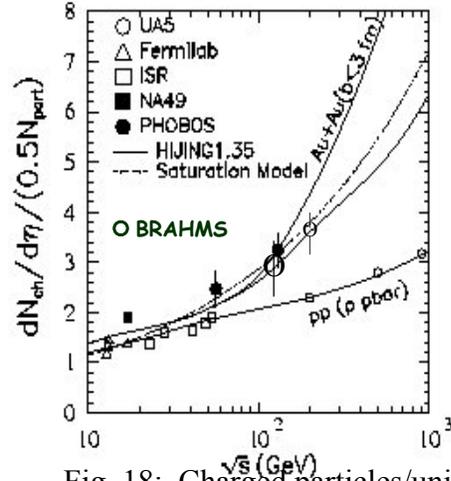

Fig. 17: Negative charged particles per unit rapidity in Au-Au collisions.

Fig. 18: Charged particles/unit rapidity vs energy.

The data on particle ratios in the central region from both the SPS and RHIC present a number of puzzles. Why is the $\bar{p}/p$ ratio significantly less than 1 as shown in Figure 19 [Jörgensen]? Are protons flowing into $\eta = 0$, even though that is 5 units of rapidity away from the target and projectile? Or are antiprotons being absorbed on the way out of the interaction region? It is also surprising that the $p/\pi$ ratio is 1 at moderate P$_T$ (Fig. 20) [d'Enterria]. This supports the hypothesis that protons from the target and projectile are making their way into the central region. Finally there is the relative increase in strange baryon production as the number of nucleons in the collision increases (Fig. 21) [Bruno].

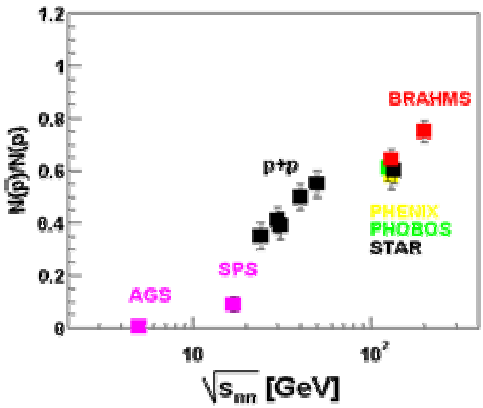
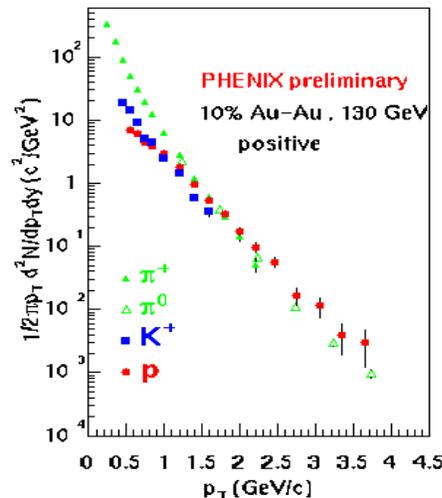

Fig. 19: The $\bar{p}/p$ ratio as a function of center-of-mass energy.

Fig. 20: Production of $\pi$, K, and p as a function of P$_T$.

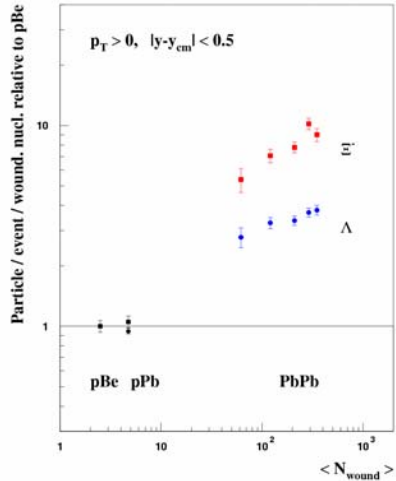 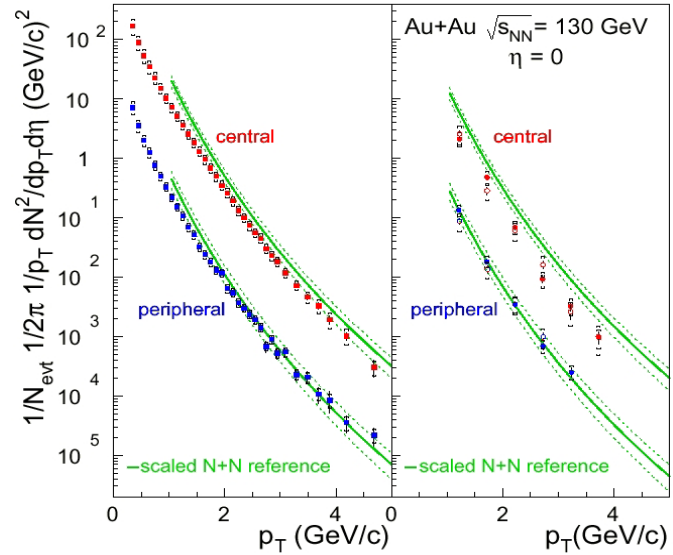

Fig. 21: Strange baryon production per wounded nucleon relative to pBe collisions from NA57.

Fig. 22: Phenix $P_T$ spectra for charged hadrons (left) and $\pi^0$'s (right).

Detailed $P_T$ spectra for produced hadrons provide important constraints on models of heavy ion collisions. For example, Figure 22 shows central production that is suppressed compared to the scaled nucleon-nucleon cross section [d'Enteria]. Tannenbaum noted in regard to such inclusive spectra that the event-by-event mean $P_T$ distribution is consistent with independent emission.

## 9. Summary

There has been a great deal of progress in understanding QCD both as an elementary interaction and as a tool needed for extracting other important physics. At this meeting, we have seen new results on hard interactions, soft processes, and collective phenomena. The future is even brighter with the prospect of much larger data samples at CLEO-c, the B factories, Tevatron Collider, RHIC, HERA, and LHC. As a result, I expect that 5 years from now, at Moriond-2007, some of today's critical questions will have been answered. However it is almost inevitable that others will take their place.

The Renconrtres de Moriond continue as a major venue for the presentation and discussion of important new scientific results, especially by young scientists. I would like to thank Jean Tran Thanh Van and the Organizing Committee for maintaining this wonderful tradition.